\documentclass[%
 reprint,showkeys,
 amsmath,amssymb,
 aps,
]{revtex4-1}

\usepackage{graphicx}% Include figure files
\usepackage{dcolumn}% Align table columns on decimal point
\usepackage{bm}% bold math
\usepackage{hyperref}% add hypertext capabilities

\usepackage{color}
\begin{document}

\preprint{APS/123-QED}

\title{Exponentially long transient time to synchronization of coupled chaotic circle maps in dense random networks}

\author{Hans Muller Mendonca}
 \affiliation{Instituto de Ci\^encias Matem\'aticas e Computa\c{c}\~ao, Universidade de S\~ao Paulo, S\~ao Carlos, S\~ao Paulo, Brazil}
\author{Ralf T\"onjes}%
%\email{toenjes@uni-potsdam.de}
\affiliation{Institute of Physics and Astronomy, Potsdam University, 14476 Potsdam-Golm, Germany}
\author{Tiago Pereira}
 \affiliation{Instituto de Ci\^encias Matem\'aticas e Computa\c{c}\~ao, Universidade de S\~ao Paulo, S\~ao Carlos, S\~ao Paulo, Brazil}

\begin{abstract}We study the transition to synchronization in large, dense networks of chaotic circle maps, where an exact solution of the mean-field dynamics in the infinite network and all-to-all coupling limit is known. In dense networks of finite size and link {probability} of smaller than one, the incoherent state is meta-stable for coupling strengths that are larger than the mean-field critical coupling. We observe chaotic transients with exponentially distributed escape times and study the scaling behavior of the mean time to synchronization.
\end{abstract}


\maketitle
\section{Introduction}

Complex nonlinear systems often exhibit collective synchronization phenomena {which can play an important role} for the overall functioning of a system~\cite{lehnertz2009synchronization,schnitzler2005normal,pikovsky2001universal}. Phase oscillator models can elucidate key aspects of the mechanism that generates the collective motion~\cite{Stankovski_RMP_2017}. The~Kuramoto model, for~instance, is particularly useful in describing groups of weakly coupled oscillators %like 
such as Josephson junctions, and~they can be analyzed in almost full detail in the thermodynamic limit of infinitely many oscillators. Indeed, Kuramoto himself initially studied the fully connected networks of coupled oscillators with frequency heterogeneity, and obtained the critical value of the coupling strength for the transition from incoherence to synchronized collective oscillations.~\cite{kuramoto1984chemical}. 

While such predictions are obtained in the thermodynamic limit, they have been used as fruitful approaches to describe networks with finitely many oscillators~\cite{bick2020understanding,eldering2021chimera}. However,  recent work has shown that finite size fluctuations or sparse connections in the network can significantly impact on the overall dynamics. In~fact, in~certain models,
synchronization cannot, even approximately, be predicted from the mean-field approximation in the thermodynamic limit~\cite{tonjes2010synchronization}. That is, in~these models, a~transition to synchronization occurs or is inhibited because of finite size fluctuations~\cite{PPikovsky2015finite,Strogatz2018}. The~interplay between mean-field predictions and finite-size fluctuations for general models remains elusive and requires further~investigation.

In this work, we study chaotic phase maps  in dense networks where the mean-field dynamics can be analyzed exactly in the thermodynamic limit. For~small coupling, due to the chaotic phase dynamics, only incoherence is stable. For~a range of coupling strengths, mean-field analysis predicts coexistence between complete chaotic synchronization and incoherence, and~for strong coupling, the incoherence becomes unstable. Then, complete synchronization is the globally attracting state in our model.
Our results are two-fold:

(i) For coupling strengths with a stable coexistence of incoherence and synchronization, although~incoherence is locally attracting, finite-size fluctuations can take the system into the basin of attraction of the absorbing state of complete synchronization. Starting near incoherence with uniformly distributed random oscillator phases, the~distribution of transient times towards synchronization is exponential and scales as a power of the system~size. 

(ii) Above the critical coupling strength, in~dense but incomplete networks, although~linear stability analysis of the mean-field equations suggests that any nonzero mean field, e.g.,~finite size fluctuations of the mean field, will grow exponentially fast, we observe an exponentially long chaotic transient in the incoherent state. Such a delayed transition to synchronization has so far not been described in dense networks of coupled phase oscillators or coupled chaotic~maps. 

\section{Model of Coupled Chaotic~Maps}

\begin{figure*}[t!]
\centering
{\includegraphics[height=4.2cm]{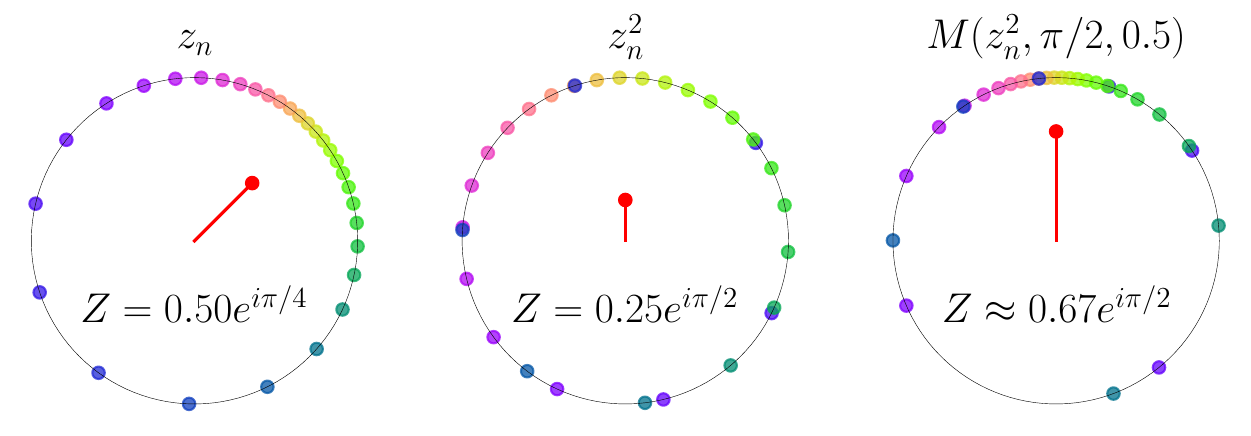}}
\caption{{\bf Dynamics of phases.} %MDPI: Please confirm if the bold should be retained.
 $N=30$ points on the complex unit circle colored by phase, and~corresponding mean field (red dot inside the unit circle). From~left to right : initial phase configuration at the points $z_n$ with mean-field amplitude $R=0.5$ and mean-field phase $\Theta=\pi/4$, after~chaotic phase doubling $z_n^2$ with $R^2=0.25$ and $2\Theta=\pi/2$, and after subsequent contraction toward the angle $\pi/2$ with intensity $\tau=0.5$.} \label{Fig:Fig01}
\end{figure*}

The local phase dynamics in each node is modelled as a Bernoulli map of the circle with time steps $t\in\mathbb{Z}$
\begin{equation}\label{Eq:Bernoulli}
\varphi(t + 1) = f(\varphi(t)) = 2\varphi(t)\mod 2\pi,
\end{equation} 
or %by 
via the abuse of notation on the complex unit circle $z = \exp(i\varphi)$, we write $z(t+1) = f(z(t)) = z(t)^2$. This map is chaotic and structurally stable~\cite{tanzi2019robustness}. That is, the~statistical properties of the map persist under small perturbations. Therefore, for~small coupling, the~maps behave as nearly independent, and~no collective dynamics is possible for small coupling. In~\cite{gong2020coupled}, the global coupling of the phase dynamics is implemented as a Moebius map
on the complex unit circle. The~Moebius map has been shown to give exact solutions of sinusoidally forced phase dynamics~\cite{MirolloStrogatz_2009Chaos}, including the Kuramoto model, Winfree-type phase equations, and %by~
via a nonlinear transformation, the~dynamics of theta neurons~\cite{PazoThetaNeurons2015}. It is therefore a meaningful alternative to the sine coupling in the standard circle map. Here, we use a composition of~\eqref{Eq:Bernoulli} and a Moebius map (see Figure~\ref{Fig:Fig01})
\begin{equation}\label{Eq:chaotic_cirlce_map}
    z{(t+1)} = M\left(f(z(t)),\Phi(t),\tau(t)\right), 
\end{equation}
where
\begin{equation}\label{Eq:Moebius}
M(w,\Phi,\tau)    = \frac{e^{i\Phi}\tau+w}{1+e^{-i\Phi}\tau w}
\end{equation}
for a coupling intensity $-1<\tau<1$, an~angle of contraction $\Phi \in \mathbb{S}^1$, and a point $w \in \mathbb{D}=\lbrace z\in\mathbb{C}:|z|<1 \rbrace$ on the open complex unit disc. The~family of Moebius maps is a group of biholomorphic automorphisms of $\mathbb{D}$, and~%by 
via analytic continuation, these transformations map the boundary of $\mathbb{D}$ bijectively onto itself. The~effect of~\eqref{Eq:Moebius} on the unit disc is a contraction of almost all points towards $\exp(i\Phi)$ on the boundary where $\lim_{\tau\to \pm 1}M(w,\Phi,\tau)=\pm\exp(i\Phi)$ and $\lim_{\tau\to 0}M(w,\Phi,\tau)=w$. The~parameter $\tau$ characterizes the strength of the contraction. For~$\tau\to 0$, the map~\eqref{Eq:chaotic_cirlce_map} approaches the uncoupled dynamics~\eqref{Eq:Bernoulli}. Moreover, the~family of wrapped Cauchy distributions
\begin{equation}\label{Eq:wrapped_cauchy}
    p(\varphi) = \frac{1}{2\pi}\frac{1-R^2}{|1-Re^{i(\varphi-\Theta)}|^2}
\end{equation}
which includes incoherence as the uniform distribution when $R\to 0$ and a delta distribution at $\varphi=\Theta$ when $R\to 1$, is invariant under~\eqref{Eq:chaotic_cirlce_map} and~\eqref{Eq:Moebius} \cite{gong2020coupled,MirolloStrogatz_2009Chaos,pikovsky2021synchronization}. This family of continuous phase measures, in~the context of phase synchronization, is known as the Ott-Antonsen manifold, and~assuming this form of phase distribution is equivalent to the so called Ott-Antonsen ansatz~\cite{ott2008low, ott2009long}. The~Ott-Antonsen manifold is parameterized %by 
using the mean-field amplitude $R$ and the mean-field angle $\Theta$
\begin{equation}
    Z = Re^{i\Theta} = \int_0^{2\pi}e^{i\varphi} p(\varphi)\,d\varphi .
\end{equation}

The mean-field amplitude $R$ is the Kuramoto order parameter~\cite{kuramoto1975self}, which is zero for incoherence, i.e.,~a uniform phase distribution, and~$R=1$ for complete synchronization $\varphi_n=\Theta$ (a.s.).
Furthermore, the~higher circular moments $Z_q$ on the Ott-Antonsen manifold with $q\in\mathbb{Z}$ are integer powers of the mean field
\begin{eqnarray}\label{Eq:circle_moms}
    Z_q = \int_{0}^{2\pi} e^{iq\varphi} p(\varphi) \,d\varphi= Z^q.
\end{eqnarray}

As a consequence, phase doubling maps the circular moments as 
$f(Z_q(t))=Z_{2q}(t) = Z_2^q(t) = f(Z_1(t))^q$, leaving the Ott-Antonsen manifold invariant and mapping the mean-field amplitude and phase as $R\to R^2$ and $\Theta\to 2\Theta$.

%\unskip

To couple the dynamics of the Bernoulli maps~\eqref{Eq:chaotic_cirlce_map}, the parameters $\Phi(t)$ and $\tau(t)$ in~\eqref{Eq:Moebius} should be defined as functions of the ensemble mean field. Following~\cite{gong2020coupled}, we define the contraction angle $\Phi(t)$ and the coupling intensity $\tau(t)$ as
\begin{eqnarray}
    Z(t) &=& \frac{1}{N}\sum_{n=1}^{N} z_n(t) = R(t)e^{i\Theta(t)}\\
    \Phi(t) &=&2\Theta(t) \\
    \tau(t) &=& \tanh\left(\frac{\varepsilon}{2}R(t) \right) \label{tau},
\end{eqnarray}
where $\varepsilon$ is a coupling strength. For~$\tau=1$, when $\varepsilon R\to \infty$, the~phases are contracted to a single point $\exp(2i\Theta)$ on the unit circle. For~small values of $\varepsilon R$, we can expand~\eqref{Eq:chaotic_cirlce_map} to the linear order and obtain the more familiar form of mean-field coupled circle maps with phase doubling
\begin{equation}
    \varphi_n(t+1) = 2\varphi_n(t) + \varepsilon R(t) \sin\left(2\Theta(t)-2\varphi_n(t)\right) + O(\varepsilon^2R^2(t)).
\end{equation}

The crucial observation is that on the Ott-Antonsen manifold, the~mean-field $Z=R\exp(i\Theta)$ transforms exactly the same way %by
via~\eqref{Eq:chaotic_cirlce_map},\eqref{Eq:Moebius} as each element $z=\exp(i\varphi)$ on the unit circle~\cite{MirolloStrogatz_2009Chaos,gong2020coupled}; that is,
\begin{equation}\label{Eq:MF_dynamics}
    Z(t+1) = M(Z^2(t),\Phi(t),\tau(t)).
\end{equation}

It is highly unusual that a closed analytic expression for the dynamics of the mean field can be derived and thus analyzed in coupled nonlinear dynamical systems. The~reduction %of 
in infinitely dimensional microscopic dynamics to the low-dimensional dynamics of the mean-field~\cite{ott2008low} has been tremendously successful in the analysis of synchronization phenomena over the last decade, while the effects of the finite system size $N$ remain difficult to \mbox{analyze~\cite{Peter2018,PeterGong2019}}. We note that the point measure of a finite ensemble of phases is never actually on the Ott-Antonsen manifold, but can, in~some sense, be arbitrarily close to the so-called thermodynamic limit, i.e.,~the limit of the infinite system size $N\to\infty$.

Applying the Ott-Antonsen ansatz to networks of phase oscillators is possible if the network structure allows for the partitioning of the vertices into a few classes of equivalent vertices. Assuming that all vertices of a class are subjected to the same sinusoidal forcing, the~dynamics of the phases in the network can be reduced to the dynamics of coupled mean fields on the Ott-Antonsen manifold for each vertex class~\cite{skardal2012hierarchical,martens2010chimeras,Strogatz2018,martens2016basins,tonjes2021coherence}. Additionally, heterogeneity in the oscillators and fluctuations in the forces can be incorporated into the mean field dynamics if they follow Cauchy distributions~\cite{laing2009chimera,tonjes2020low,clusella2022regular}. 

\subsection{Mean-Field~Analysis}

The mean-field dynamics~\eqref{Eq:MF_dynamics} can be written in terms of the polar representation
\begin{equation}\label{Eq:OA_MF_RTh}
\Theta(t+1) = f(\Theta(t)) \,\,\,\,\,  \mbox{~and }  \,\,\,\,\, R(t+1) =\frac{\tau(t)+R^2(t)}{1+\tau(t) R^2(t)}.
\end{equation}

This means that the dynamics of the phase $\Theta$ decouples from the amplitude and will evolve chaotically. Using Equations~(\ref{tau}) and (\ref{Eq:OA_MF_RTh}), we obtain the amplitude dynamics
\begin{equation}\label{Eq:R}
R(t+1) = \frac{\tanh\left(\frac{1}{2}\varepsilon R(t)\right) +R^{2}(t)}{1+\tanh\left(\frac{1}{2}\varepsilon R(t)\right) R^2(t)}
\end{equation}
which describes the exact evolution of the order parameter $R$ in a closed form. We can readily determine the fixed points of the mean-field amplitude $R(t)$ and their linear stability. Both the complete synchronization $R=1$ and the complete desynchronization $R=0$ are fixed points of~\eqref{Eq:R}, and change stability at unique critical points $\varepsilon_{1}=\ln(2)\approx 0.69$ and $\varepsilon_{0}=2$, respectively, as~determined by the eigenvalues of Jacobian of Equation~(\ref{Eq:R}) at these fixed points. These critical points are connected by an unstable fixed point branch $(\varepsilon(R_u),R_u)$, where
\begin{equation}
    \varepsilon(R_u) = \frac{1}{R_u}\log\left(\frac{(1+R_u)^2}{1+R_u^2}\right).
\end{equation}

\begin{figure}[t!]
%\centering
\setlength{\unitlength}{1cm}
\begin{picture}(4.2,4.2)
\put(0,0){\includegraphics[height=4.2cm]{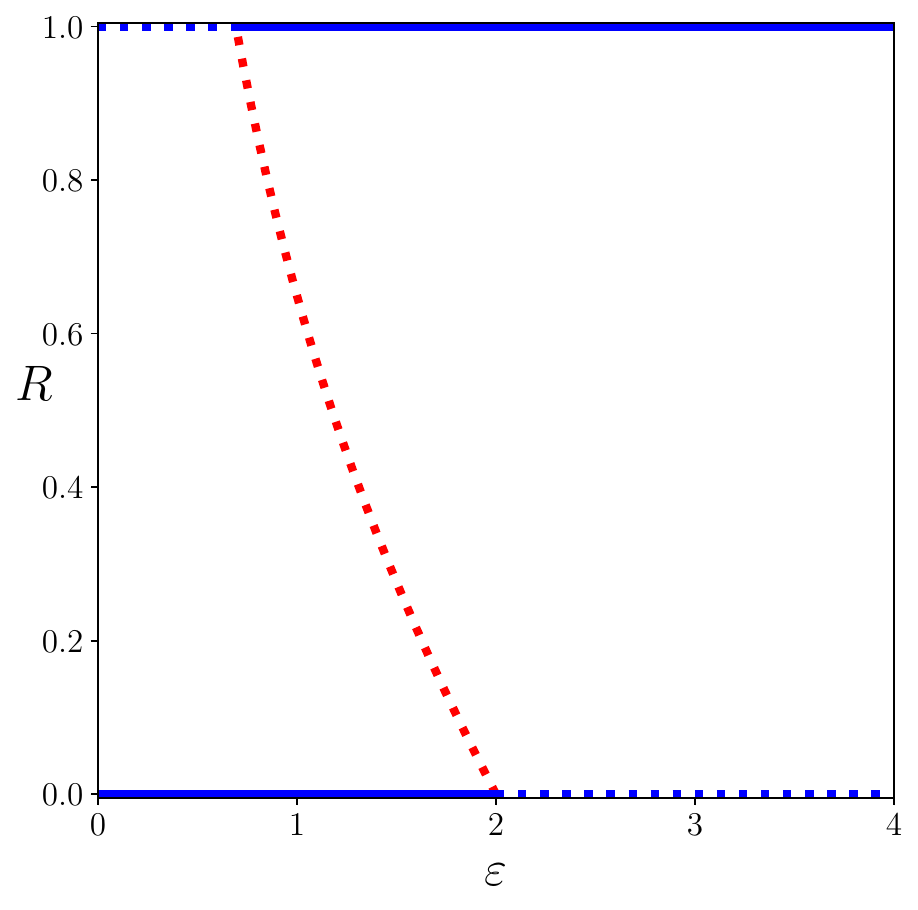}}
\put(-0.2,3.8){(\textbf{a})}
\end{picture}
\begin{picture}(4.2,3.8)
\put(0,0.49){\includegraphics[height=3.6cm]{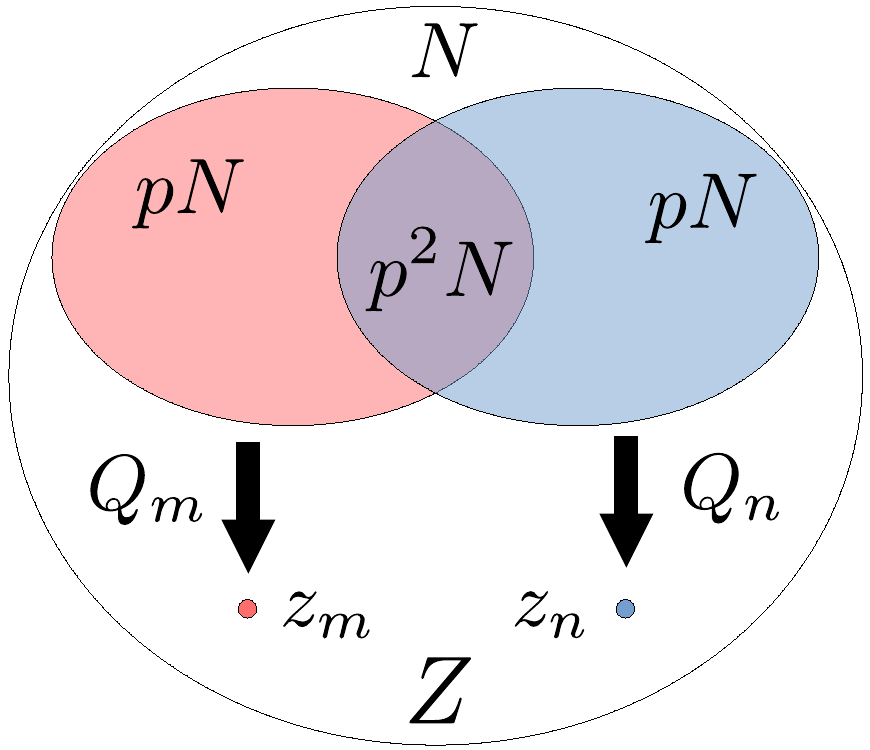}}
\put(0.0,3.8){({\bf b})}
\end{picture}
\caption{{\bf Bifurcation diagram of the mean-field amplitude and a representation of the network interaction}. In~(\textbf{a}), the bifurcation diagram of the all-to-all coupling mean-field dynamics~\eqref{Eq:OA_MF_RTh}, i.e.,~on the Ott-Antonsen manifold. Dotted lines show linearly unstable fixed points and solid lines show linearly stable fixed points in the thermodynamic limit. (\textbf{b}) Venn diagram of a dense network with $N$ vertices and connection probability $p$. The~sets of neighbors of nodes $m$ and $n$ are of size $pN$ and their overlap is of size $p^2N$, resulting in correlated local mean fields $Q_m=R_m\exp(i\Theta_m)$ and $Q_n=R_n\exp(i\Theta_n)$ acting on the states $z_m$ and $z_n$. The~ratio of the amplitudes of the local mean fields and the global mean field are independent of the network size $N$.} \label{Fig:Fig02}
\end{figure}

This expression is derived from~\eqref{Eq:R} by setting $R(t+1)=R(t)=R_u$ and resolving the equation for $\varepsilon$.
This means that this system of all-to-all coupled, identical chaotic phase maps will always evolve to complete synchronization or complete desynchronization, with~a small region $\ln(2)< \varepsilon<2$ of bistability (Figure \ref{Fig:Fig02}a).

\subsection{Extension to~Networks}

Next, we have studied the same phase dynamics on a random network of $N$ maps which are coupled to exactly $k$ different, random neighbors. Here, each phase $\varphi_n$  couples to a local mean field
\begin{equation}\label{Eq:localMF}
    Q_n = 
    R_n e^{i\Theta_n} = \frac{1}{k} \sum_{n=1}^N A_{nm}z_m
\end{equation}
where $A_{nm}$ are the entries of the adjacency matrix, i.e.,~equal to one if there is a link from vertex  $m$ to vertex  $n$, but zero otherwise, and $k = \sum_{m=1}^N A_{nm}$ is the in-degree of node  $n$, which, for computational simplicity, we assume to be identical for all nodes. Thus, with $\tau_n = \tanh\left(\frac{\varepsilon}{2}R_n\right)$, the dynamics of the phases coupled through a network are
\begin{equation}
    z_n(t+1) = \frac{e^{2i\Theta_n(t)}\tau_n(t)+z_n^2(t)}{1+e^{-2i\Theta_n(t)}\tau_n(t) z_n^2(t)}.
\end{equation}

A class of networks is dense if $\lim_{N\to\infty} \langle k\rangle/N = p > 0$, where $\langle k\rangle $ is the mean node degree. Therefore, $p$ is 
the fraction of nodes, in~relation to the system size $N$, that an oscillator is coupled to. Since dense networks are defined in the limit of $N\to\infty$, there is no sharp distinction between sparse and dense networks of finite size. We refer to a finite network as dense if~two nodes share more than one neighbor on average, i.e.,~$\langle k\rangle ^2/N=p^2N>1$. In~large dense networks, the~local mean fields of the oscillators in the neighborhood of each node~\eqref{Eq:localMF} are equal to the global mean field, with a deviation of $O(1/\sqrt{k})$, where $k$ is the size of the neighborhood, i.e.,~the in-degree of the node. Therefore, mean-field theory should be exact for dense networks in the thermodynamic limit where $\langle k \rangle \to\infty$.

{\it  The network model} %MDPI: Please confirm if the italics should be retained.}. 
First, we wish to compare the simulation results directly with our mean-field analysis. For~large random networks with a link density $p=k/N$ and $0<p<1$, the~numerical simulations are time-consuming since the $N$ local mean fields at each node in the network need to be computed in each time step. To~simplify these computations, we use a random network where each node couples to exactly $k$ different random neighbors. This model with a unique in-degree of $k$ for each node is slightly different from the Erd\"os Renyi model, with a Poissonian in-degree distribution of small relative width $\textrm{std}(k)/\langle k \rangle \sim 1/\sqrt{k}$. For~large $k$, the results of the simulations in our random network model and other random networks with uncorrelated node degrees and a vanishing relative width of the degree distribution are expected to be~identical.

\section{Results}
\unskip
\subsection{Distributions of Transient~Times}

We perform a large number $M$ of simulations $m=1\ldots M$ from independent, uniformly distributed random initial phases over a maximum of $T$ steps and record in each simulation the first time step $t_m$ when $R\ge 0.5$, i.e.,~the transition time from an incoherent state to  complete synchronization. Finite-size scaling for such a discontinuous transition is challenging~\cite{de2015generic}. 
The exponential distribution of the times $t_m$, according to some characteristic transition rate, can  be checked in a rank plot of time points $t_m$, which gives the sample complementary cumulative distribution $C(t)=\textrm{prob}(t \ge t_m)=\textrm{rank}(t_m)/M$ %MDPI: Figures should be cited in numerical order. Thus, "Ref 4" should be cited after "3".
 \ref{Fig:Fig04}a,d). 

An exponential tail distribution $C(t)$ up to observation time $T$ indicates an exponential distribution of transient times. Since the simulation time is finite, transition times $t_m\ge T$ are not observed, which represents a problem when we are interested in the average time to synchronization. However, assuming a discrete exponential, i.e.,~geometric distribution, a~maximum likelihood estimation of the average transition time is possible up to values considerably exceeding the observation time $T$ (see Appendix \ref{app1}). 

Denoting the number of simulations that synchronize at times $t_m<T$ as $M_{T}$, and~defining the observable values $l_m=\textrm{min}(t_m,T)$, the~maximum likelihood estimation of the expected value $T_{esc}=\textrm{E}[t_m]$ for the geometric distribution is
\begin{equation}\label{Eq:MLE_Tesc}
    T_{esc} = \frac{\left\langle l_m\right\rangle M}{M_T}.
\end{equation}
with the sample mean $\langle l_m\rangle$. If~the transition to synchronization is observed in all simulations, i.e.,~$M_T=M$, the~estimator is simply the sample mean of $t_m$, which is an estimator of $T_{esc}$ for arbitrary transient time distributions. However, when most runs do not synchronize within the finite simulation time $T$, the~ratio $M/M_T$ contains additional information, and~the estimated mean escape time can be much larger than the observation~time.

\subsection{During Coexistence: Escape over the Unstable~Branch}
In~\cite{luccioli2012collective}, it was reported that the transition from incoherence to collective dynamics in sparse networks of coupled logistic maps is of the mean-field type. The~analysis in~\cite{restrepo2005onset} predicts a shift in the critical coupling strength in random networks of Kuramoto phase oscillators of the order $\langle k\rangle^2/\langle k^2\rangle$ due to degree inhomogeneity, and $1/\langle k\rangle$ due to finite size fluctuations of the local mean fields. That is, in~dense, homogeneous networks with $\langle k\rangle^2/\langle k^2\rangle\to 1$ and $\langle k\rangle\to\infty$, the critical coupling strength does not change.
We expected to find similar behaviors for network-coupled Bernoulli maps. 
In complete or almost complete networks $k/N=p\approx 1$ for $\varepsilon < 2$, there is a small probability that finite size fluctuations bring the order parameter $R$ above the unstable branch, leading to a spontaneous transition to complete synchronization, as shown in Figure~\ref{Fig:Fig03}a. %We found that the transient time to synchronization is exponentially distributed and depends strongly on the network size $N$, as~shown in Figure~\ref{Fig:Fig04}a-?).
We first observe the scaling of the transient time in fully connected networks with $p=1$.
For values of $\varepsilon<\varepsilon_0=2.0$, the~transition rate to synchronization scales strongly with the size $N$ of the system \mbox{(Figure \ref{Fig:Fig04}b,c)}. However, for~values $\varepsilon>\varepsilon_0$, the average transition time depends very weakly on $N$, as~the system grows exponentially fast from a state of incoherence, with $R\approx 1/\sqrt{N}$. We estimate a finite size scaling exponent $\beta$ below the transition threshold by collapsing the curves $T_{esc}(\varepsilon,N)$ using the ansatz
$
    T_{esc}(\varepsilon,N) = T_{esc}\left((\varepsilon-\varepsilon_0)N^\beta\right).
$
The data %is 
are consistent with an ad~hoc exponent of $\beta = 1/3$ (Figure \ref{Fig:Fig04}c).

\subsection{Above the Critical Coupling Strength: Long Chaotic~Transient} 
\begin{figure}[t!]
%\centering
\setlength{\unitlength}{1cm}
\begin{picture}(4.2,4.2)
\put(0,0){\includegraphics[height=4.1cm]{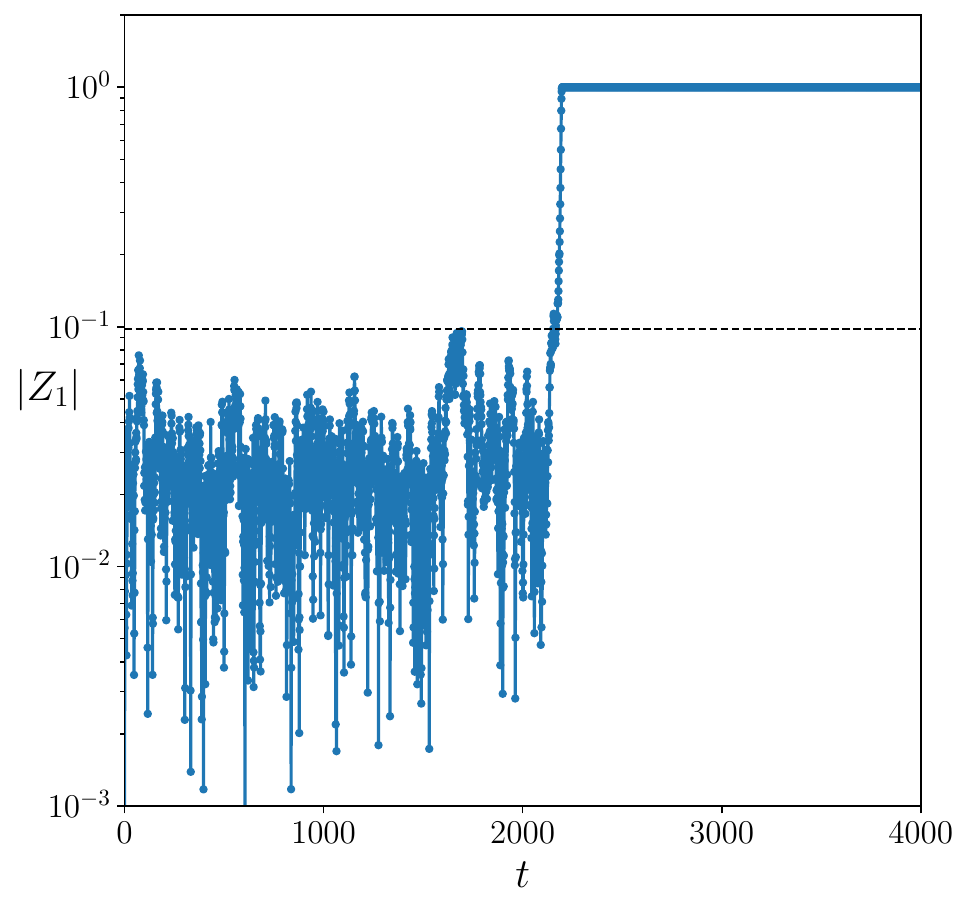}}
\put(-0.2,3.8){(\textbf{a})}
\end{picture}
\begin{picture}(4.2,4.2)
\put(0,0){\includegraphics[height=4.1cm]{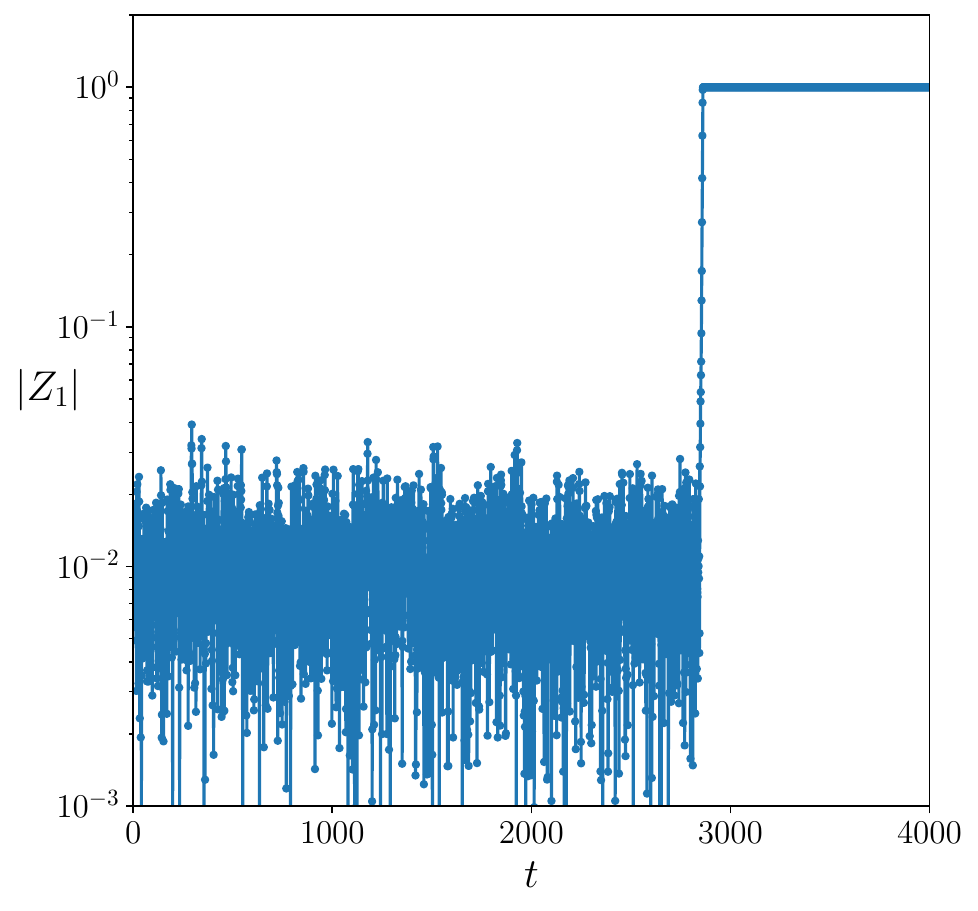}}
\put(0,3.8){(\textbf{c})}
\end{picture}
\\
\begin{picture}(4.2,4.2)
\put(0,0){\includegraphics[height=4.1cm]{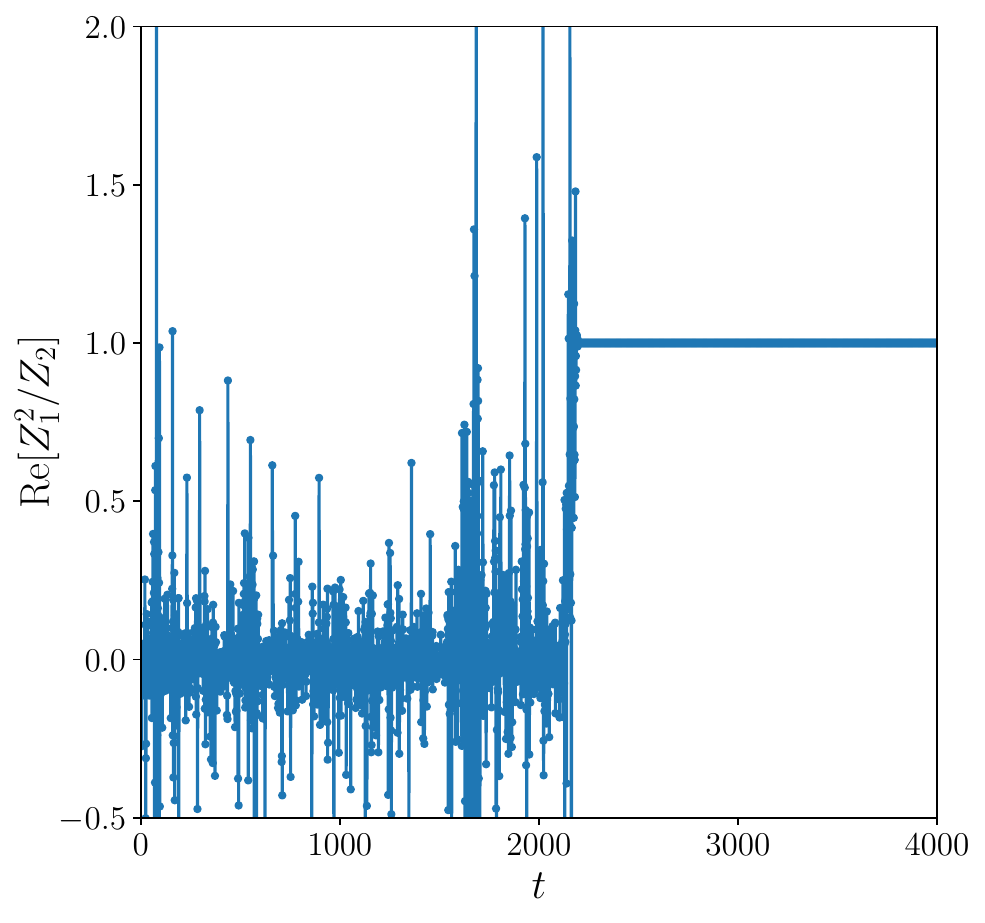}}
\put(0,3.8){(\textbf{b})}
\end{picture}
\begin{picture}(4.2,4.2)
\put(0,0){\includegraphics[height=4.1cm]{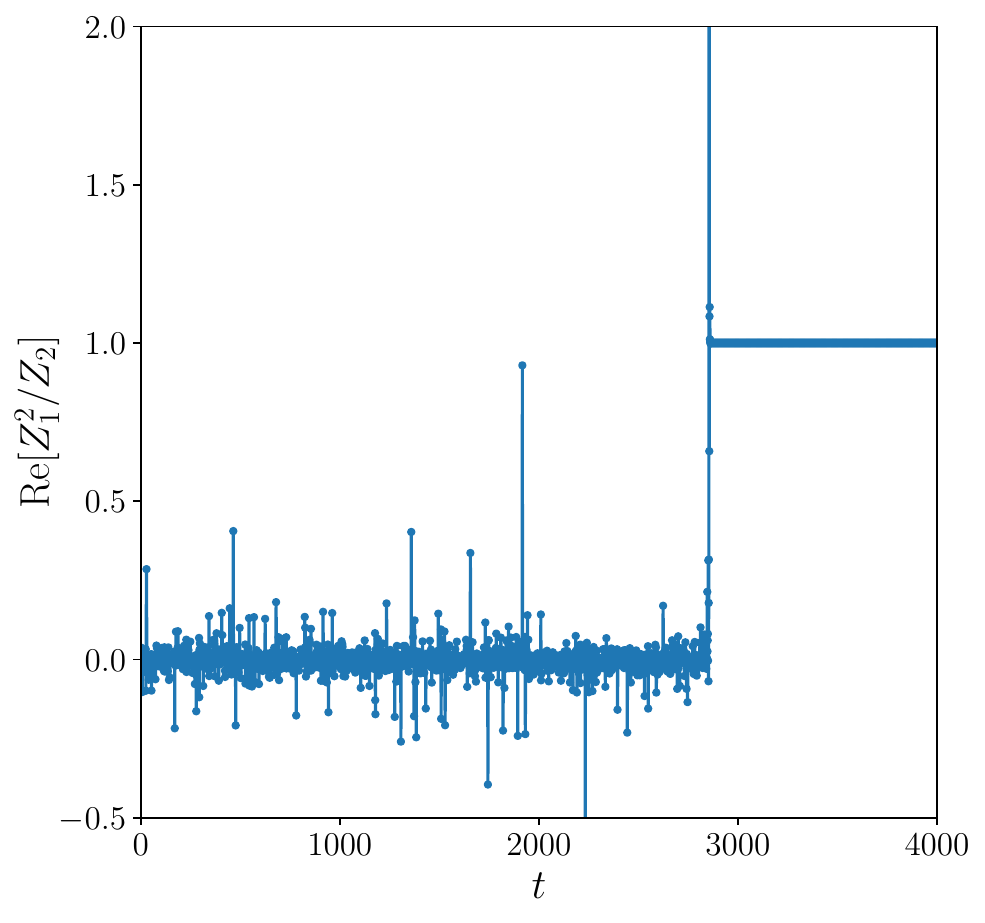}}
\put(0,3.8){(\textbf{d})}
\end{picture}
\caption{{\bf Transient to synchronization} for $N=$ 10,000 coupled maps in (\textbf{a},\textbf{b}), a fully connected network with coupling strength $\varepsilon=1.81$ below the critical coupling $\varepsilon_0=2$, and (\textbf{c},\textbf{d}), in a random network with connection probability $p=0.1$ for a coupling strength of $\varepsilon=2.3$ above the critical coupling. The~upper panels (\textbf{a},\textbf{c}) show the order parameter $R(t)$, and the lower panels (\textbf{b},\textbf{d}), the real part of the ratio of the first two circular moments $\textrm{Re}[Z_1^2/Z_2]$. This serves as a visual measure of the alignment of the system state with the Ott-Antonsen manifold, where the ratio is exactly equal to one. The~dashed line in (\textbf{a}) marks the value of the unstable fixed point of the mean-field dynamics, $R_u=0.098$. Above~that value, the~state of complete synchronization is attractive on the Ott-Antonsen manifold. In~(\textbf{b},\textbf{d}), the incoherent state $R=0$ is unstable; however, finite size fluctuations do not grow exponentially. Instead, we observe a long chaotic~transient.} \label{Fig:Fig03}
\end{figure}
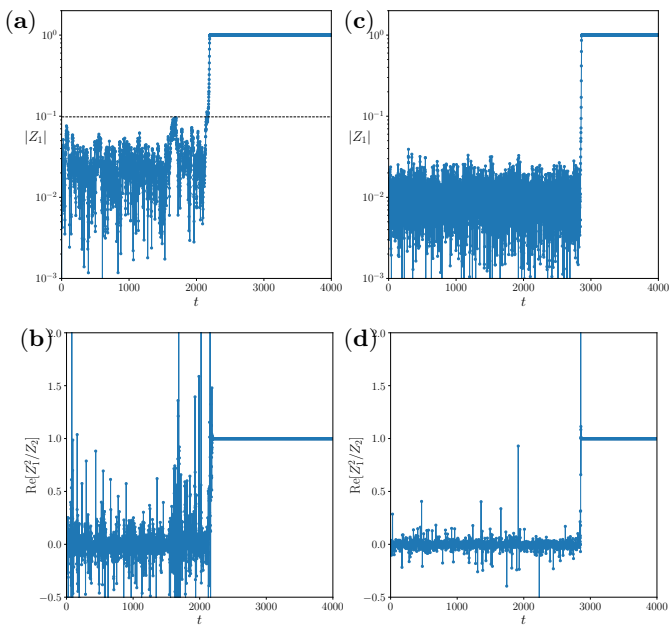

Above the critical coupling strength $\varepsilon>\varepsilon_0=2$, we expected finite size fluctuations to grow exponentially fast and independently of $N$, as~predicted by linear stability analysis of the mean-field equations~\eqref{Eq:R}. Instead, for~small connection probabilities $0<p<1$, we have observed a chaotic transient with seemingly stationary finite size fluctuations $O(1/\sqrt{N})$ of the mean field (Figure \ref{Fig:Fig03}). In~the large $N$ limit, the~distribution of the transient times depends on the link density $p$ with increasingly long transients as $p$ is decreased, but it is otherwise independent of $N$. 

A coupling strength for which a transition to complete synchronization could still be observed within the simulation time was considerably larger than the mean-field critical coupling $\varepsilon_{0}=2$.  That is, even in dense networks and above the mean-field critical coupling, finite size fluctuations will not necessarily result in the nucleation and exponential growth of a collective mode.  Such a delayed transition to synchronization~\cite{baer1989slow} has so far not been described in systems of coupled phase oscillators~\cite{ichinomiya2004frequency,restrepo2005onset,ko2008partially} or coupled logistic maps~\cite{luccioli2012collective}.

In Figure~\ref{Fig:Fig04}f, we plot $T_{esc}$ over $(\varepsilon-\varepsilon_0)p$ to demonstrate that the average transition time is roughly scaling as $1/p$. We do not look for higher-order corrections such as a weak dependence of $\varepsilon_0$ on $p$, although~the curves do not collapse perfectly. Note that the escape time is largely independent of the network size (Figure \ref{Fig:Fig04}e,f). For~$p=0.1$, $0.05$, and $0.025$ we have performed  simulations with $N=10^4$ (circles) and with $N=5\times 10^4$ (crosses) for comparison. For~$p=0.01$, we compare network sizes $N=10^4$ (circles) with very time-consuming simulations in networks with $N=10^5$ (crosses).

\unskip

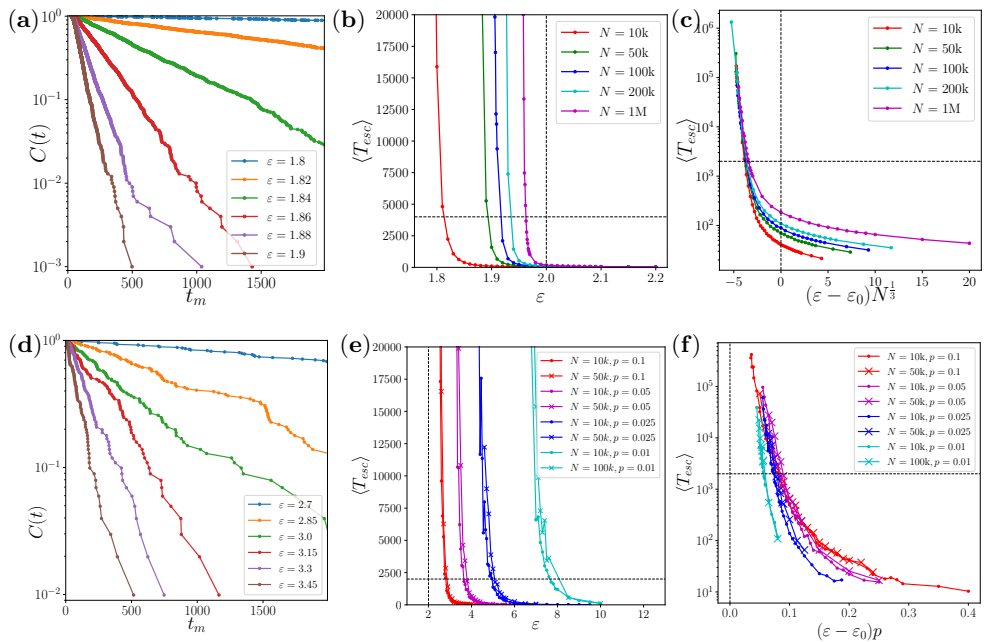
\begin{figure*}[t!]
\centering
\setlength{\unitlength}{1cm}
\begin{picture}(4.2,4.2)
\put(0,0){\includegraphics[height=4.1cm]{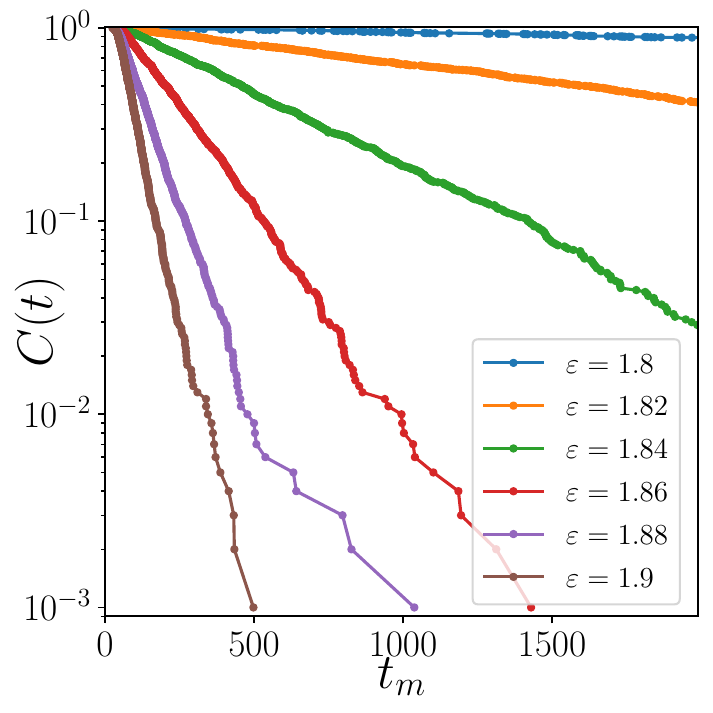}}
\put(-0.2,3.8){(\textbf{a})}
\end{picture}
\begin{picture}(4.2,4.3)
\put(0,0){\includegraphics[height=4.1cm]{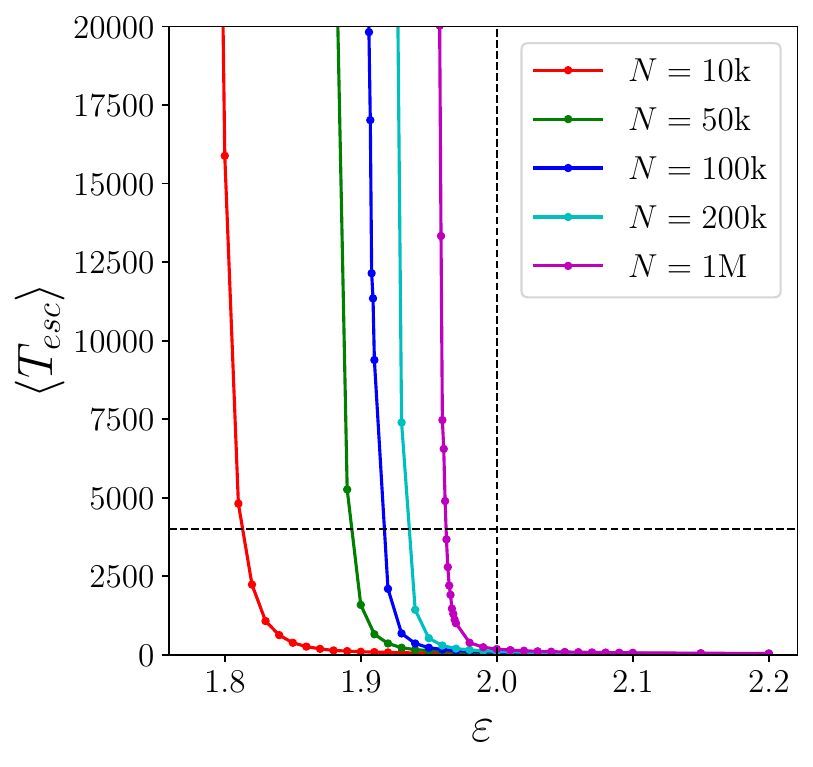}}
\put(-.2,3.8){(\textbf{b})}
\end{picture}
\begin{picture}(4.2,4.2)
\put(0.,0){\includegraphics[height=4.1cm]{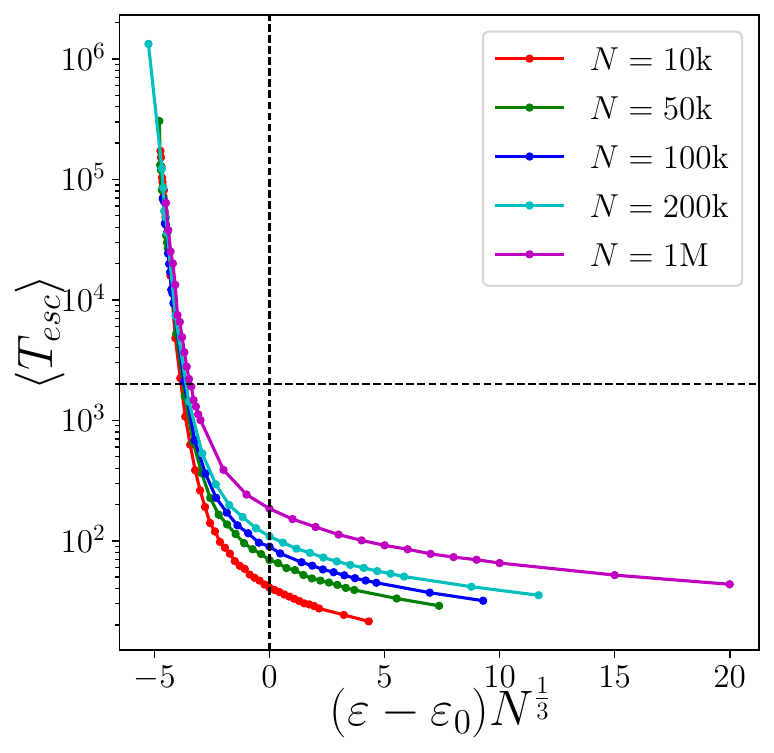}}
\put(0,3.8){(\textbf{c})}
\end{picture}\\
\begin{picture}(4.2,4.2)
\put(0,0){\includegraphics[height=4.1cm]{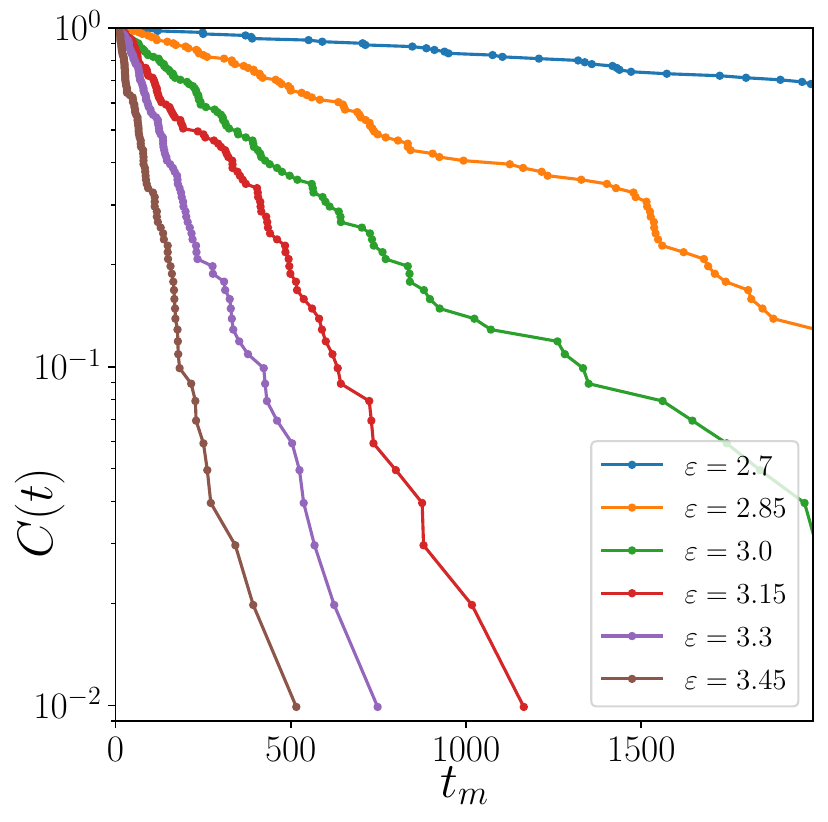}}
\put(-0.3,3.8){ (\textbf{d}) }
\end{picture}
\begin{picture}(4.2,4.3)
\put(0,0){\includegraphics[height=4.cm]{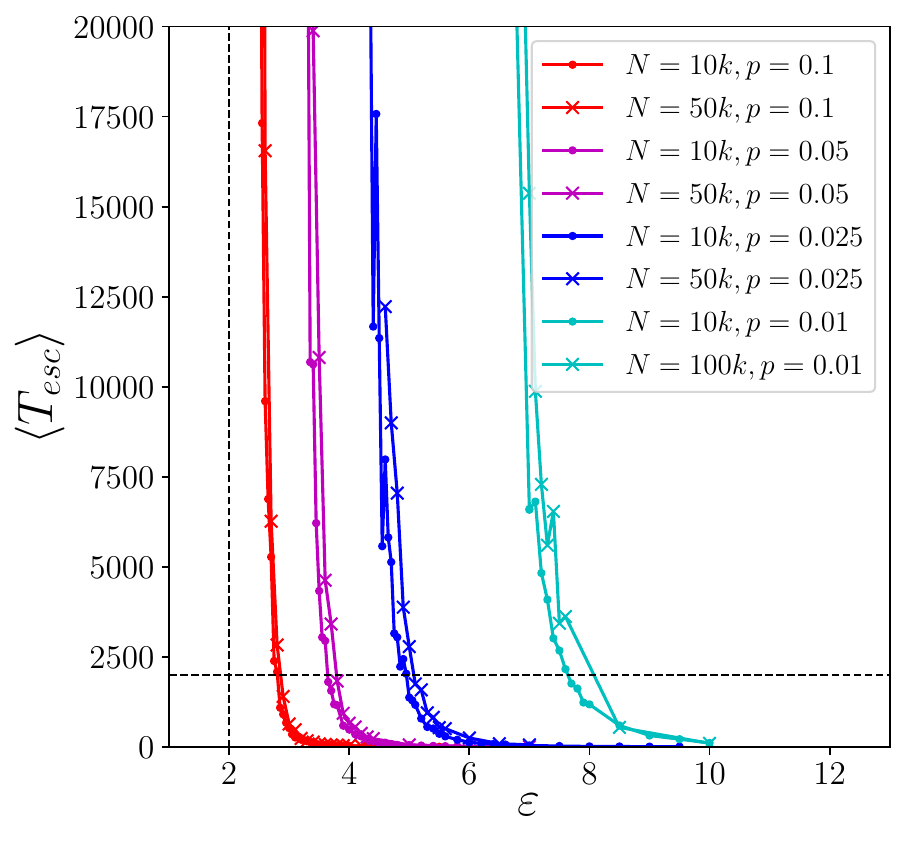}}
\put(-0.2,3.8){ (\textbf{e}) }
\end{picture}
\begin{picture}(4.2,4.2)
\put(0,-.1){\includegraphics[height=4.1cm]{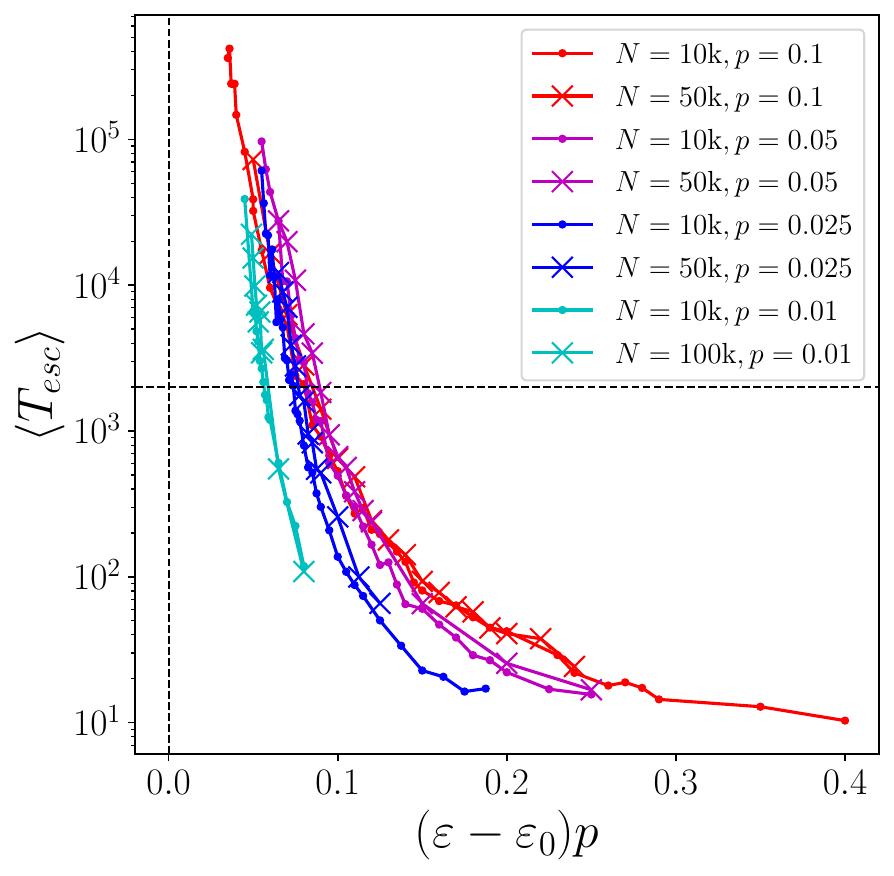}}
\put(-0.1,3.8){ (\textbf{f}) }
\end{picture}
\caption{{\bf Statistics %MDPI: Please use commas to separate thousands for numbers with five or more digits (not for four digits) in the picture. e.g., "10000" should be "10,000"
 of transient times $t_m$ to synchronization}. (\textbf{a}--\textbf{c}) In the fully connected network; %and 
 (\textbf{d}--\textbf{f}) in random networks of various link densities $p=k/N$. The~left panels show straight lines in semi-logarithmic plots of cumulative tail distributions of the transient times, demonstrating the rate character of the transition process. The~middle panels show the estimated average transient times for various combinations of system sizes $N$, coupling strengths $\varepsilon$, and~link densities $p$. The~mean field critical coupling strength $\varepsilon_0=2.0$ and the maximum observation time $T$ are marked by dashed lines. In~the globally coupled system in pannels (\textbf{a}--\textbf{c}), the~transient time depends strongly on the system size $N$, whereas in dense networks and above $\varepsilon_0$ (\textbf{d}--\textbf{f}), the transient time depends strongly on the link density $p=k/N$, but not on the system size. We demonstrate the scaling of the transient times in panels (\textbf{c}) and (\textbf{f}) on the right. In~the globally coupled system, the~exponential divergence of the transient times below $\varepsilon_0$ appears to be a function of $(\varepsilon-\varepsilon_0)N^\frac{1}{3}$. In~dense networks, the~exponential divergence is roughly a function of $(\varepsilon-\varepsilon_0)p$.}
 \end{figure*}

\subsection{Discussion of Finite Size~Scaling}

Mean field theory assumes a phase distribution on the Ott-Antonsen manifold. The~characteristic function of a wrapped Cauchy distribution is the geometric sequence $Z_q=Z^q$ of circular moments~\eqref{Eq:circle_moms}.
However, in~the incoherent state with $N$ independent uniformly distributed phases $\varphi_n$ the circular moments of an ensemble
\begin{equation}
    Z_q = \frac{1}{N}\sum_{n=1}^N e^{iq\varphi_n}
\end{equation}
are almost independent complex numbers with a Gaussian distribution of mean zero and a variance of $1/N$ by virtue of the central limit theorem. The~action of the Bernoulli map on the circular moments is the shift
\begin{equation}\label{Eq:DoubleZ}
    Z_q\to Z_{2q},
\end{equation}
that is, it is achieved by discarding all odd circular moments.
The exponential growth of the order parameter in accordance to mean field theory is expected after the distribution comes close to the Ott-Antonsen manifold, i.e.,~when the first few circular moments align by chance sufficiently under the mapping 
\eqref{Eq:DoubleZ}; in~particular, $Z_2(t)\approx Z_1^2(t)$. Unless~the directions of $Z_2$ and $Z_1^2$ align by chance, as~they would on the Ott-Antonsen manifold, the~subsequent contraction of strength $\varepsilon R$ in the direction of $Z_1^2$ after the phase doubling may even decrease the amplitude of the order parameter.
In addition, for~coupling strengths $\varepsilon$ below the critical value, $R=|Z_1|$ must be above the unstable branch $R>R_u(\varepsilon)\sim (\varepsilon_0-\varepsilon)$. 

The rate of such a random event should depend on the ratio between $R_u(\varepsilon)$ and the standard deviation $1/\sqrt{N}$ of the Gaussian distribution of the complex mean field. Based on this scaling argument, the~expected time to synchronize should scale as\linebreak $T_{esc}=T_{esc}((\varepsilon-\varepsilon_0)\sqrt{N})$ below the critical coupling. The~best collapse of the estimated escape times in fully connected networks of coupled Bernoulli maps was observed by scaling the distance to $\varepsilon_0$ with $N^{1/3}$ (Figure \ref{Fig:Fig03}c), i.e.,~the exponential divergence of the escape time approaches $\varepsilon_0$ slower than $1/\sqrt{N}$ in the thermodynamic limit. One possibility for this discrepancy is that the scaling argument only considers the chance of $R>R_u$ and not the alignment process of the higher-order circular~moments.

Above the critical coupling strength, there is only the condition of the alignment of circular moments with the Ott-Antonsen manifold for the initiation of exponential growth. Since in the incoherent state, all circular moments are random Gaussian with identical variance, the~alignment process~\eqref{Eq:DoubleZ} is strictly independent of the system size $N$. Once exponential growth in the direction of the Ott-Antonsen manifold occurs, the~time to synchronization is logarithmic, that is, it is weakly dependent on $N$. However, it appears that the alignment with the Ott-Antonsen manifold needs to be stronger for networks with link densities of $p<1$. For~small link densities, the~divergence of the escape time occurs at larger values $\varepsilon>\varepsilon_0$. This is reminiscent of stabilization by noise~\cite{khasminskii2011stochastic}, where a system is driven away from a low-dimensional unstable manifold of a fixed point into stronger attracting stable~directions. 

In simulations of dense random networks of coupled Bernoulli maps, we could see the independence of the mean escape time from the network size and the scaling of the escape time with roughly $\sim 1/p$ (Figure~\ref{Fig:Fig04}f). To~explain this scaling, we argue that mean field theory might be extended to dense networks, where each node couples to a finite neighborhood of $pN$ nodes in the network, and~for every two nodes, these neighborhoods overlap on a set of size $p^2 N$ (Figure \ref{Fig:Fig02}b). The~local mean fields are Gaussian random forces of mean value $Z$, variance $1/k=1/pN$, and a pairwise correlation of $p$, which is the~relative size of the overlap. The~decrease in correlation between the local mean fields in networks with link densities $p<1$ can be interpreted as individual, finite size noise on the maps, which couple to the global mean field, plus some uncorrelated random deviation. Therefore, the~contractions of the phases do not occur in the same direction for different nodes in the network. The~strength of the contraction in the direction of the mean field is effectively reduced by the factor $p$,  i.e., 
\begin{eqnarray}
    \tau = \tanh\left(\frac{1}{2}\varepsilon R\right) p \approx \frac{1}{2}\varepsilon p R
\end{eqnarray}
shifting the coupling strength dependence of the transition time (above $\varepsilon_0$) by a factor of $1/p$.

\section{Conclusions}

We have investigated the synchronization of coupled chaotic maps in dense random networks, utilizing mean-field equations and examining network configurations with different link probabilities.
Firstly, we noticed the existence of chaotic transients to synchronization within these networks. This means that the incoherent state can persist for extended periods before transitioning into synchronization. This finding led us to study the statistics of transient times and their scaling behaviors in the process of synchronization. The~transition times follow exponential distributions, indicating spontaneous transitions at a constant rate. It is noteworthy that the transition from incoherence to complete synchronization only occurs spontaneously in networks of finite size.
Additionally, we have observed a remarkable dependence of the transient times to synchronization on the link probability $p$, represented by the ratio of the in-degree to the total number of nodes, at~coupling strengths where an immediate transition to synchrony would be expected from mean field theory. Whether such a delayed transition is due to the specifics of our model or is typical for a more general class of dynamics remains an open~question.

%%%%%%%%%%%%%%%%%%%%%%%%%%%%%%%%%%%%%%%%%%
%\section{Conclusions}%
%
%This section is not mandatory, but can be added to the %manuscript if the discussion is unusually long or complex.
%
%%%%%%%%%%%%%%%%%%%%%%%%%%%%%%%%%%%%%%%%%%
%\section{Patents}
%
%This section is not mandatory, but may be added if there are patents resulting from the work reported in this manuscript.
%
%%%%%%%%%%%%%%%%%%%%%%%%%%%%%%%%%%%%%%%%%%
\vspace{6pt} 

%%%%%%%%%%%%%%%%%%%%%%%%%%%%%%%%%%%%%%%%%%
%% optional
%\supplementary{The following supporting information can be downloaded at:  \linksupplementary{s1}, Figure S1: title; Table S1: title; Video S1: title.}


%%%%%%%%%%%%%%%%%%%%%%%%%%%%%%%%%%%%%%%%%%

This research was funded by the FAPESP CEMEAI 391, Grant No. 2013/07375-0, Serrapilheira Institute (Grant No.Serra-392 1709-16124), Newton Advanced Fellow of the Royal Society
%English Editor: Please check that meaning is preserved below
(393 NAF$\backslash$R1$\backslash$180236), CAPES and CNPq, Grant No 166191/2018-3.

%%%%%%%%%%%%%%%%%%%%%%%%%%%%%%%%%%%%%%%%%%
%% Optional
\section{Apendix}
Here, we calculate the maximum likelihood estimation for the mean value of a geometric distribution $P(t;\alpha)=(1-\alpha)\alpha^t$ for discrete values $t=0,1,\ldots$ of time steps when only times $t<T$ can be observed. The~expected value for the geometric distribution is
\begin{equation}\label{Eq:geometric_mean}
    \textrm{E}\left[ t \right] = (1-\alpha)\sum_{t=0}^\infty t \alpha^t = \frac{\alpha}{1-\alpha}.
\end{equation}
Since the times $t_m$, $m=1\ldots M$ are only observable up to step $T-1$, we define $l_m = \textrm{min}(t_m,T)$. The~probabilities for the possible values of $l_m$ are
\begin{eqnarray}
    P(l_m=T;\alpha) &=& 1-(1-\alpha)\sum_{t=0}^{T-1}\alpha^t = \alpha^T \label{Eq:MLE_probs_T} \\
    P(l_m=t<T;\alpha) &=& (1-\alpha)\alpha^t. \label{Eq:MLE_probs_n}
\end{eqnarray}
\\
The derivative of the log-likelihood of $M$ independent observations $l_m$ with respect to the parameter $\alpha$ is
\begin{equation}\label{Eq:logLH_derivative}
    \frac{\partial_\alpha P(l_1,l_2,\ldots,l_M;\alpha)}{P(l_1,l_2,\ldots,l_M;\alpha)} = \sum_{m=1}^M \frac{\partial_\alpha P(l_m;\alpha)}{P(l_m;\alpha)}.
\end{equation}
For the probabilities~\eqref{Eq:MLE_probs_T},\eqref{Eq:MLE_probs_n}, the derivatives are
\begin{eqnarray}
    \frac{\partial_\alpha P(l_m=T,\alpha)}{P(l_m=T,\alpha)} &=& \frac{T}{\alpha} \label{Eq:logLH_diff_T}\\
    \frac{\partial_\alpha P(l_m=t<T,\alpha)}{P(l_m=t<T,\alpha)} &=& \frac{t}{\alpha}-\frac{1}{1-\alpha}. \label{Eq:logLH_diff_n}
\end{eqnarray}
For a maximum of the log-likelihood for the observed values $l_m$, the sum in~\eqref{Eq:logLH_derivative} is required to be zero. Inserting $M-M_T$ times the term~\eqref{Eq:logLH_diff_T} for all observations $l_m=T$ and $M_T$ terms~\eqref{Eq:logLH_diff_n}, one  for each observation $l_m=t<T$, we obtain
\begin{equation}\label{Eq:MLE_condition}
    (M-M_T)\frac{T}{\alpha} + \sum_{l_m<T} \frac{l_m}{\alpha} - M_T\frac{1}{1-\alpha} = 0.
\end{equation}
With
\begin{equation}
    \langle l_m\rangle = \frac{1}{M}\sum_{m=1}^M l_m = \frac{1}{M}\left((M-M_T)T + \sum_{l_m<T}l_m\right)
\end{equation}
we can divide~\eqref{Eq:MLE_condition} by the number $M$ of observations and re-order the equation to obtain
\begin{equation}
    \frac{\langle l_m \rangle M}{M_T} = \frac{\alpha}{1-\alpha}.
\end{equation}
However, this is exactly the expected value $\textrm{E}\left[ t \right]$ of time steps for the full geometric distribution~\eqref{Eq:geometric_mean}. %The maximum likelihood estimation of the expected value for a continuous exponential distribution $p(x)=\alpha e^{-\alpha x}$ is exactly the same, $T_{esc} = \langle l_m \rangle M / M_{T}$ where $l_m = \textrm{min}(t_m,T)$ but requires a likelihood function with a mixed continuous $(0\le l_m< T)$ and discrete $(l_m=T)$ distribution.

\bibliography{chaotic_map_sync}
\end{document}